\documentclass[11pt,letterpaper]{article}
\pdfoutput=1
\usepackage{natbib}
\usepackage{graphicx} 
\usepackage{tabularx,booktabs}
\usepackage{url}
\usepackage{fullpage}
\usepackage{lipsum}
\newcommand{\junk}[1]{}
\usepackage[linesnumbered]{algorithm2e}
\usepackage{mathtools} 
\usepackage{listings} 
\usepackage{arydshln} 
\usepackage{threeparttable} 
\usepackage{float}

\title{MirLibSpark: A Scalable NGS Plant MicroRNA Prediction Pipeline for Multi-Library Functional Annotation}
\author
{
\centering
Chao-Jung Wu
\and Amine M. Remita
\and Abdoulaye Banir\'e Diallo
\footnote{Corresponding author}
\\Department of Computer Science, Universit\'e du Qu\'ebec \`a Montr\'eal,\\ Montr\'eal, H3C 3P8 Canada\\
}
\date{May 2019}

\newcommand\blfootnote[1]{%
  \begingroup
  \renewcommand\thefootnote{}\footnote{#1}%
  \addtocounter{footnote}{-1}%
  \endgroup
}

\begin{document}
\maketitle

\blfootnote{Authors' emails: wu.chao\_jung@courrier.uqam.ca, remita.amine@courrier.uqam.ca, diallo.abdoulaye@uqam.ca}
\blfootnote{\textsuperscript{\textcopyright} Chao-Jung Wu, Amine M. Remita, Abdoulaye Baniré Diallo, 2019. This is the author's version of the work. It is posted here for your personal use. Not for redistribution. The definitive version was published in the \textit{Proceedings of the 10th ACM International Conference on Bioinformatics, Computational Biology and Health Informatics} (ACM-BCB) pp. 669-674, 2019, https://doi.org/10.1145/3307339.3343463}

\begin{abstract}
The emergence of the Next Generation Sequencing increases drastically the volume of transcriptomic data. 
Although many standalone algorithms and workflows for novel microRNA (miRNA) prediction have been proposed, 
few are designed for processing large volume of sequence data from large genomes, 
and even fewer further annotate functional miRNAs by analyzing multiple libraries.
We propose an improved pipeline for a high volume data facility by implementing mirLibSpark based on the Apache Spark framework.
This pipeline is the fastest actual method, and provides an accuracy improvement compared to the standard.
In this paper, we deliver the first distributed functional miRNA predictor as a standalone and fully automated package. 
It is an efficient and accurate miRNA predictor with functional insight. 
Furthermore, it compiles with the gold-standard requirement on plant miRNA predictions.
\end{abstract}

\section{Introduction}
The emergence of the Next Generation Sequencing (NGS) allows researchers to quickly and cheaply collect billions of expressed small RNAs (sRNAs) sequences according to a given condition so that the volume of small-RNAome data increases dramatically.
However, the annotation of sRNAs has never been rapidly and systematically achieved.
Parallelization approach is becmoing useful for computation-intensive tasks in biology \citep{biegert2006mpi}.
In the era of big data, CPU speeds and storage capacity are greatly improving.
Hard disk input/output (I/O) is also dealt with by Hadoop MapReduce.
Furthermore, Resilient Distributed Dataset (RDD) abstraction is performed to imporve Hadoop's disk-based computing \citep{zaharia2012resilient}.
Scalable MapReduce methods are revolutionizing traditional bioinformatics tools in RNA research \citep{mckenna2010genome,nellore2016rail,de2017sparkblast}.

Nowadays, several well-known bioinformatics tools are dedicated to annotating sRNAs as miRNAs in plants.
MiRDeep-P \citep{yang2011mirdeep} employs a probabilistic scoring algorithm to predict novel miRNAs from precursor signatures and annotates directly known miRNA candidates as miRNAs without considering their dynamic expressions within the transcriptome.
Currently, the best prediction is performed by ShortStack \citep{axtell2018revisiting,axtell2013shortstack}, and the fastest prediction tool is miR-PREFeR \citep{lei2014mir}.
Both ShortStack and miR-PREFeR incorporate miRNA annotation criteria \citep{meyers2008criteria} in their annotation algorithms and decrease false positive predictions so that they annotate miRNAs with low false positives, high precision and high accuracy \citep{axtell2018revisiting}.
Although ShortStack and miR-PREFeR take multiple pre-aligned sRNA-seq libraries as inputs, ShortStack is not scalable and has difficulty to deal with species having larger genomes such as maize that has 2 GB genome \citep{lei2014mir}.
MiR-PREFeR spawns subprocesses to execute a function on distributed input data across multiple processors on a given machine.
This approach improved the processing speed for maize sRNA-seq libraries \citep{lei2014mir}.
However, the performance of any parallelization approach has not been assessed in species that have large genomes, such as wheat of 13 GB genome.

In our previous work \citep{agharbaoui2015integrative}, we established a standardized integrative procedure to annotate miRNAs.
The procedure includes four major steps: miRNA prediction, target gene prediction, differential expression analysis and functional enrichment.
This data-intensive annotation procedure reduces false positive predictions by taking into account experts criteria, and renders meaningful miRNA annotations due to the incorporation of functional analysis. However, none of the existing tools has worked towards scaling up the entire procedure.
To the best of our knowledge, there is only one miRNA predictor employing MapReduce \citep{pan2015design} to perform a few steps of the procedure. Additionally, MR-microT server \citep{kanellos2014mr} uses MapReduce to annotate the downstream target genes of miRNAs.

In this work, we propose an improved miRNA prediction procedure that adds the miRNA annotation criteria for big data \citep{axtell2018revisiting}, incorporates functional and structural annotations and accelerates the processing time.
We present mirLibSpark, a Spark-based miRNA annotation pipeline that executes the entire miRNA annotation procedure and is able to process multiple sRNA-seq libraries with high prediction performance and speed in both personal computer and high performance computing clusters.

\section{MirLibSpark architecture and execution}

\subsection{MirLibSpark overview}

\begin{center}
\begin{figure}[H]
\centering
\includegraphics[width=8.5cm]{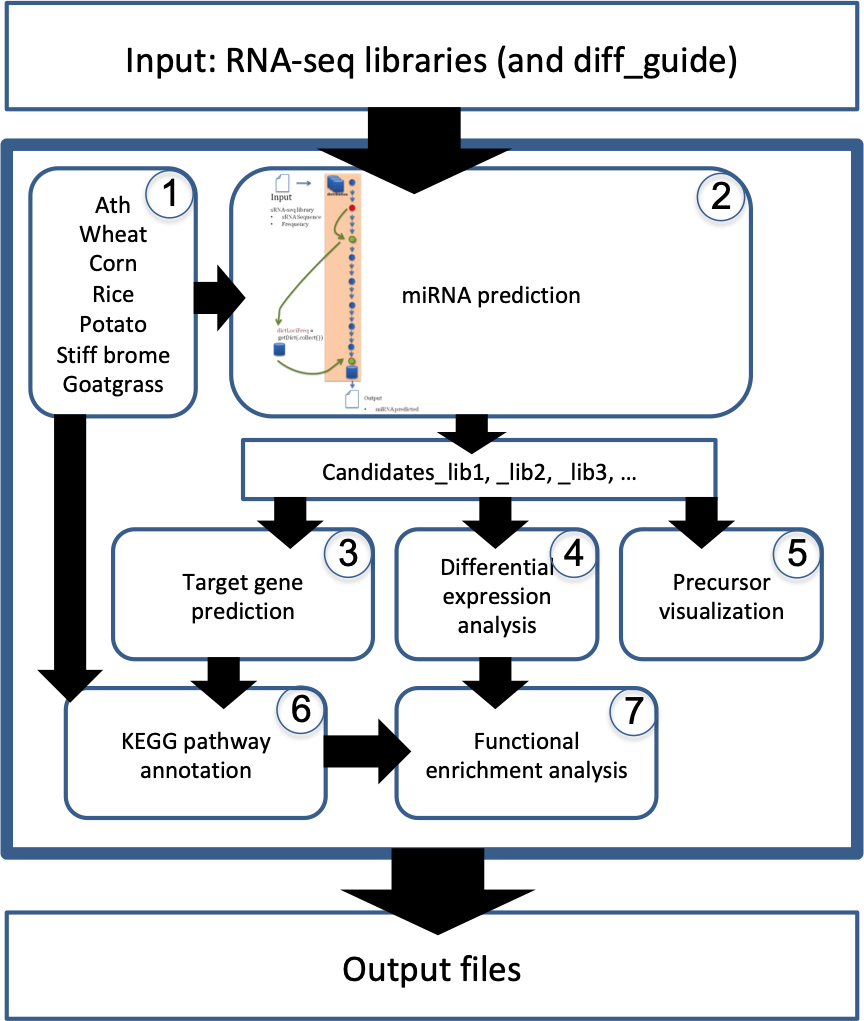}
\caption{MirLibSpark overview. M1: genomic reference annotations. M2: miRNA prediction. M3: target gene prediction. M4: differential expression analysis. M5: precursor visualization. M6: KEGG pathway annotation. M7: functional enrichment analysis. M for module.}
\label{FIG::overview}
\end{figure}
\end{center}

\textbf{Annotation modules.}
MirLibSpark contains modules dedicated to integrative procedure of functional miRNA annotation, as shown in Figure \ref{FIG::overview}.
It includes a module (M1) providing the reference genomes, Bowtie indices \citep{langmead2009ultrafast} and KEGG pathway functional annotations \citep{kanehisa2017kegg}. 
Its core module is the miRNA prediction module (M2). 
Additional modules are attached to help users to better annotate miRNAs (M3: target gene prediction, M6: KEGG pathway annotation of the target genes and M5: precursor visualization using VARNA \citep{darty2009varna}) and to aggregate multi-library data (M4: differential expression analysis and M7: functional enrichment analysis).

\textit{MiRNA prediction module (M2).} The mirLibSpark pipeline implements the four major computational steps for miRNA prediction: filtering, alignment, folding and prediction. For each run, mirLibSpark pipeline is initiated with a tunable configuration, 
then 30 different operations are applied on the multiple RDD created objects to make resilient and distributed data.
These operations apply rigorously the miRNA annotation criteria and procedures for plants defined in \citep{meyers2008criteria,axtell2018revisiting}.
The parameters were already optimized and were designed to be adjustable in the pipeline, because the experts' criteria may evolve in the future.

\textit{Additional functional annotation modules (M3-M7).} Post-prediction processing among multiple libraries is performed in one run.
We compute the statistics of differential expression to select functionally relevant miRNA candidates \citep{benjamini1995controlling,kal1999dynamics}.
Their target genes are subsequently analyzed for pathway enrichment \citep{vlachos2015diana,kanehisa2017kegg}.

\textbf{Implementation.}
Based on the computing load, computation steps within miRNA prediction module (shown in Figure \ref{FIG::1combo}), precursor visualization module and target gene prediction module are chained through Spark RDD operations \citep{zaharia2012resilient}.
In each step, an RDD object is generated to distribute the instances and operations across a cluster of computers. 
Each instance corresponds to a sRNA sequence and its related data.
We used \textit{map}() and \textit{filter}() functions when it is required, where \textit{map}() changes the content of an instance and \textit{filter}() eliminates an instance based on a boolean statement.

We opt for the implementation of adapted functions to the Spark architecture for data and computing distribution instead of using extensively external programs. 
These programs should require less dependency and be currently maintained and open source to facilitate the pipeline redistribution.
All the external programs are either distributed with the project or are automatically installed by mirLibSpark.
Therefore, installation procedures for these programs are not required by the user.

\textbf{Availability.}
MirLibSpark is available as an open source project in GitHub and as a docker image, along with its manual, in the following link
https://github.com/bioinfoUQAM/mirLibSpark.\\
We deliver mirLibSpark as a standalone package for two environments, (1) a personal computer or (2) a remote server with high computational power.
For biologists, we recommend to use the docker image of mirLibSpark in their personal computers.
For bioinformaticians, the GitHub project can be deployed in any server.
Upon installation of mirLibSpark, it is ready to execute experiments for a selected genome in a personal computer or with minimal configuration in a high performance computing server.

\subsection{MiRNA prediction module (M2)}
\begin{center}
\begin{figure}[h]
\centering
\includegraphics[width=10cm]{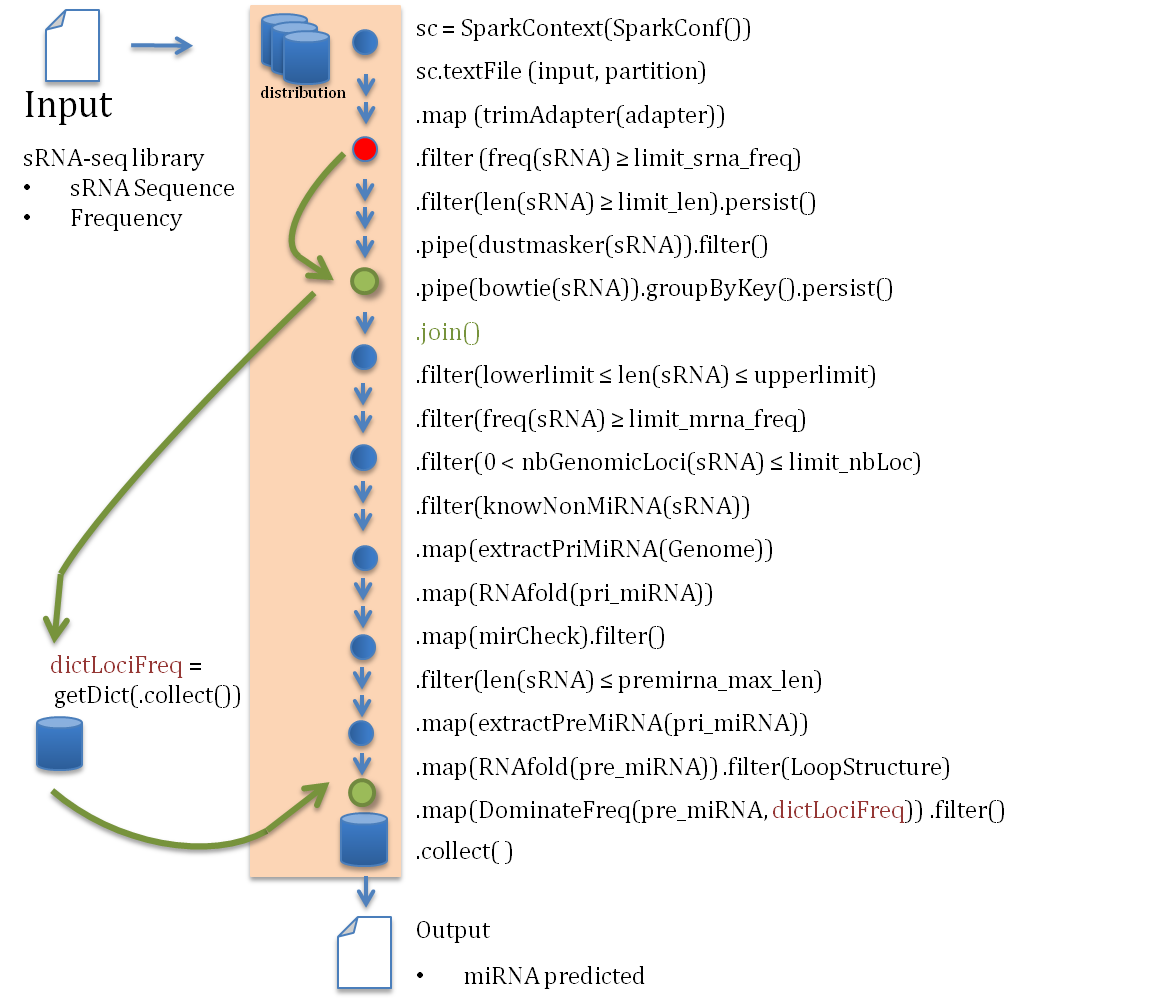}
\caption{MirLibSpark miRNA prediction module.}
\label{FIG::1combo}
\end{figure}
\end{center}

A simplified illustration of the RDD chaining in the prediction module is shown in Figure \ref{FIG::1combo}.
A $SparkContext()$ is initiated with adjustable parameters of Spark and mirLibSpark.
Default values, which are optimized for plants, are used if not otherwise specified by users.
In each submission request, the pipeline can handle one to several sRNA-seq libraries.
The first RDD object is created by reading the data of each input library.  
Thereafter, operations are applied on RDDs until the output is obtained and printed to a file.
Several filters, including low complexity filter using dustmasker \citep{morgulis2006fast} and various miRNA and precursor length and abundance criteria \citep{meyers2008criteria,axtell2018revisiting}, are achieved by the $filter()$ operation.
Only aligned sequences using Bowtie within a small number of genomic loci are retained.
The RDD generated by alignment is reused as reference for the filter step of the miRNA candidate expression within precursor range.
After mapping, only the sRNAs with length between 21 and 24 nucleotides (nt) are retained \citep{axtell2018revisiting}.
A set of known non-miRNA sequences according to the annotation of the genome are removed.
Pre-miRNA candidates surrounding the sRNA genomic loci are extracted and their secondary structures are computed using RNAfold \citep{Lorenz2011}. 
The RNA structures and the decisions of miRCheck \citep{jones2004,meyers2008criteria} are rendered by $map()$ functions with specified rules as parameters.
The precursors longer than 300 nt or containing a second loop that is larger than 5 nt are discarded \citep{axtell2018revisiting}.
Finally, a dominantly expressed potential miRNA within the pre-miRNA candidate sequence is considered a true miRNA and is subjected to subsequent functional analysis.
Most of the parameters are easily adjustable as needed.
Since the parameters are optimized for plants and can affect the efficiency and the accuracy of the pipeline, users do not need to change them unless suggested by the experts.

Several incorporated programs, such as Bowtie, already support parallelization, and indeed are not a time-consuming step.
Therefore, the acceleration by our implementation is to parallelize and chain every single prediction step in one run.
For example, RNAfold is parallelized within mirLibSpark, and giving mirCheck the resulting structure as input is also taken care of by mirLibSpark.

\subsection{Additional annotation modules}
\textbf{Genomic annotation (M1).}
To enhance user's experience, mirLibSpark already supports seven plant species of small and large genomes:
\textit{Arabidopsis thaliana} (ath), \textit{Triticum aestivum} (wheat), \textit{Zea mays} (corn), \textit{Oryza sativa} (rice), \textit{Solanum tuberosum} (potato), \textit{Brachypodium distachyon} (stiff brome) and \textit{Aegilops tauschii} (Tausch's goatgrass).
The project includes a script for user to fetch the genomes and annotations from Ensembl plants (http://plants.ensembl.org/) \citep{bolser2017ensembl} and to build automatically Bowtie indices for supported plant species. 
The genome-wide annotations of KEGG pathways are retrieved from KEGG database \citep{kanehisa2017kegg}.

\textbf{Target gene prediction (M3).}
The best 100 target genes are predicted by miRanda \citep{john2004human}, parallelized in mirLibSpark.
Gene variants are consolidated to a single identifier. 
Then the best 5 scored identifiers are subjected for functional analysis.

\textbf{Differential expression analysis (M4).}
According to the guide file, comparisons are performed between libraries.
Expression fold change for each miRNA candidate is calculated.
Then Kal's test \citep{kal1999dynamics} is performed to examine the difference between samples by calculating a \textit{z}-score and a \textit{p}-value for each miRNA.
The \textit{p}-value is corrected by false discovery rate \citep{benjamini1995controlling}.
Only the miRNAs with fold changes greater than 2 or less than 0.5, and corrected \textit{p}-values smaller than 0.05 are considered differentially expressed.

\textbf{KEGG pathway annotation (M6) and functional enrichment analysis (M7).}
Target genes of differentially expressed miRNAs in the libraries are annotated with KEGG pathways by KEGG pathway annotation module (M6).
Pathway enrichment is performed by hypergeometric analysis, similar to other miRNA functional analysis software mentioned in \citep{vlachos2015diana}.
The chance of all the drawing combinations of the genes to obtain the sample frequency of a function based on the background frequency is the \textit{p}-value.
Functions with \textit{p}-values smaller than 0.05 are considered enriched.

\subsection{Execution}

\textbf{Configuration options.}
Tunable parameters and their default values are accessible in mirLibSpark, including the minimum frequency of sRNA and miRNA, minimum length of sRNA, length range of miRNA, maximum length of pre-miRNA, maximum number of genomic loci, precursor searching range, extra flank length to the validated precursor for folding visualization, temperature used for RNAfold, 
and parameter options for miRCheck, miRanda and Spark.
Several parameters are related to the choice of species, such as genome, index, target file and pathway annotation.
If a supported species is chosen, these related parameters are automatically set accordingly.
In the case of custom species, users will provide the files and update the parameters.

\textbf{Inputs.}
MirLibSpark takes as input one to several sRNA-seq libraries of the same input format.
Four input formats are supported: tsv (readcount), reads, FASTA and FASTQ.
A guide file (\textit{diffguide\_file}) defining the control-experiment pair is required if functional analysis modules are activated for a multi-library task.
As an example, for a task analyzing two libraries (Lib1 and Lib2), if the user wishes to define that Lib1 is the control for Lib2, the guide file has to be provided by the user and formatted as follows.
\begin{verbatim}
Experiment->Control
Lib2->Lib1
\end{verbatim}

\textbf{Outputs.}
For any request, it would generate the first 10 of the output files as listed in the manual.
For a request processing more than two libraries,
if differential expression analysis parameter (DF) is set true, then two more outputs are produced;
if KEGG pathway enrichment analysis parameter is set true, with the condition that DF is true, then one more output is produced.

\section{Preparations for Benchmarking}
\subsection{Construction of simulated data sets}

\textit{Positive data set.}
The positive data set (PD) is the entire high confidence \textit{Arabidopsis thaliana} miRNAs of miRBase version 21 \cite{kozomara2018mirbase}.
The same miRNA sequences may have different precursors and are registered as different miRNAs.
Only unique miRNA sequences were collected for this study.
In version 21, there are 349 unique miRNA registries for \textit{Arabidopsis}, and a subset of them are high confidence (100).

\textit{Negative data set.}
The negative data set (ND) was constructed as follows.
The universe of non-miRNA is presumably larger than true miRNA.
Therefore, the number of ND sequences is 1,000 by applying an arbitrary factor of 10 on the number of PD.
The ND sequences were generated by extracting sequences from \textit{Arabidopsis} genome at random positions with lengths between 18 and 25 nt (inclusive).
Sequences found in miRBase v21 \textit{Arabidopsis} miRNAs were excluded in the ND collection.
Sequences annotated as CDS, rRNA, snoRNA, snRNA and tRNA in TAIR10 were also excluded in the ND collection.

\subsection{sRNA-seq libraries}
We evaluated the behavior and the performance of mirLibSpark with real sRNA-seq libraries.
GSE44622 series (27 libraries in total) \citep{jeong2013comprehensive}, GSM518432, GSM738727 \citep{lei2014mir} and GSM1087974 \textit{Arabidopsis} sRNA-seq libraries were obtained from NCBI Gene Expression Omnibus (GEO) repository.
The \textit{Triticum aestivum} sRNA-seq library, S002B35, was obtained from Li et al \citep{li2018transcriptomic}.

\subsection{Implementation of a sequential version}
The python script, $sequential.py$, is a prediction pipeline that executes in memory the major prediction steps as in the mirLibSpark but without Spark data management or parallelism. 

\subsection{Implementation of a miRDeep-P wrapper}
The miRDeep-P predictor is semi-automated and requires users to execute several scripts and commands one after another. 
We implemented the python script, $run\_mirdeep\_p.py$, to execute the prediction of known and novel miRNAs automatically with the same parameters in mirLibSpark.

\subsection{miRNA clustering}
Similar sequences in the predicted miRNA set are clustered as a group based on the bitscore (greater than 20) given by blastn alignment in blastn-short task mode \citep{altschul1997gapped}.
Bitscore indicates a log2-scaled estimation of the search space that is required to find a score as good as or better than this one by chance.
When the bitscore is 20, we would have to search about $2^{20} = $ one million sequences to find a sequence similar to our query by chance.

\section{Benchmarking and Discussions}
\subsection{Improved execution time}
The performance of several plant miRNA predictors is compared with mirLibSpark (shown in Table \ref{TAB::compare_software}).
In the simulated \textit{Arabidopsis} data sets, we incorporated 100 miRBase v21 high confidence miRNAs as positive data set. 
We extracted 1,000 unannotated random short sequences from the genome of \textit{Arabidopsis} to construct the negative data set.
Execution time and prediction performance of mirLibSpark and the other tools were assessed by the processing of the simulated data sets with specified number of cores.
For the real sRNA-seq NGS data, GSN518432 was used \citep{lei2014mir}.
MirLibSpark incorporates several miRNA structure and expression criteria \citep{meyers2008criteria,axtell2018revisiting}. 
Therefore, it has better F1 score and accuracy measure than ShortStack on the simulated data sets.
Its processing of the \textit{Arabidopsis} sRNA-seq library is 3.6 times faster than that of miR-PREFeR.

\begin{table}[h!]
\centering
\begin{threeparttable}
\caption{Performance of plant miRNA annotation software using (A) simulation data, (B) real world data adapted from \citep{lei2014mir}}
\label{TAB::compare_software}
\begin{tabular}{l|rrrr}
\hline
(A)  & mirLibSpark & \begin{tabular}[c]{@{}l@{}}miR-\\Deep-P\end{tabular} & ShortStack &\\ \hline
nbCore & 32 & 1 & 1 & \\
Time (sec) & 135 & 194 & 389 & \\
TP & 84 & 69 & 22 & \\
FP & 9 & 57 & 0 & \\
F1 & 0.87 & 0.61 & 0.36 & \\
MCC & 0.86 & 0.57 & 0.45 & \\
\end{tabular}
\begin{tabular}{l|rrrrrr}
\hline\hline
(B)*  & mirLibSpark & ShortStack & mirPrefer & mirDeepP \\\hline
nbCore         & 32          & 1          & 32        & 1        \\
Time (sec)     & 120         & 2191       & 577       & 74601    \\
TP             & 39          & 31         & 56        & 47       \\
TN             & 757514      & 757575     & 757247    & 757469   \\
FP             & 54          & 1          & 304       & 91       \\
FN             & 151         & 159        & 134       & 143      \\
F1             & 0.276       & 0.279      & 0.204     & 0.287    \\
MCC            & 0.293       & 0.398      & 0.214     & 0.29     \\
\hline
\end{tabular}
\begin{tablenotes}
    \small
    \item nbCore: number of cores allocated. Time: execution time in seconds. TP: true positives. FP: false positives. F1: F-measure. MCC: Matthews correlation coefficient. 
     *: Used the standard method to define positive and negative sets in the library, where negative set were the sequences in the library not registered in miRBase, and thus the FP predictions could be novel miRNAs.
\end{tablenotes}
\end{threeparttable}
\end{table}

\subsection{Comparison of performance with other predictors}
\begin{center}
\begin{figure}[h]
\centering
\includegraphics[width=12cm]{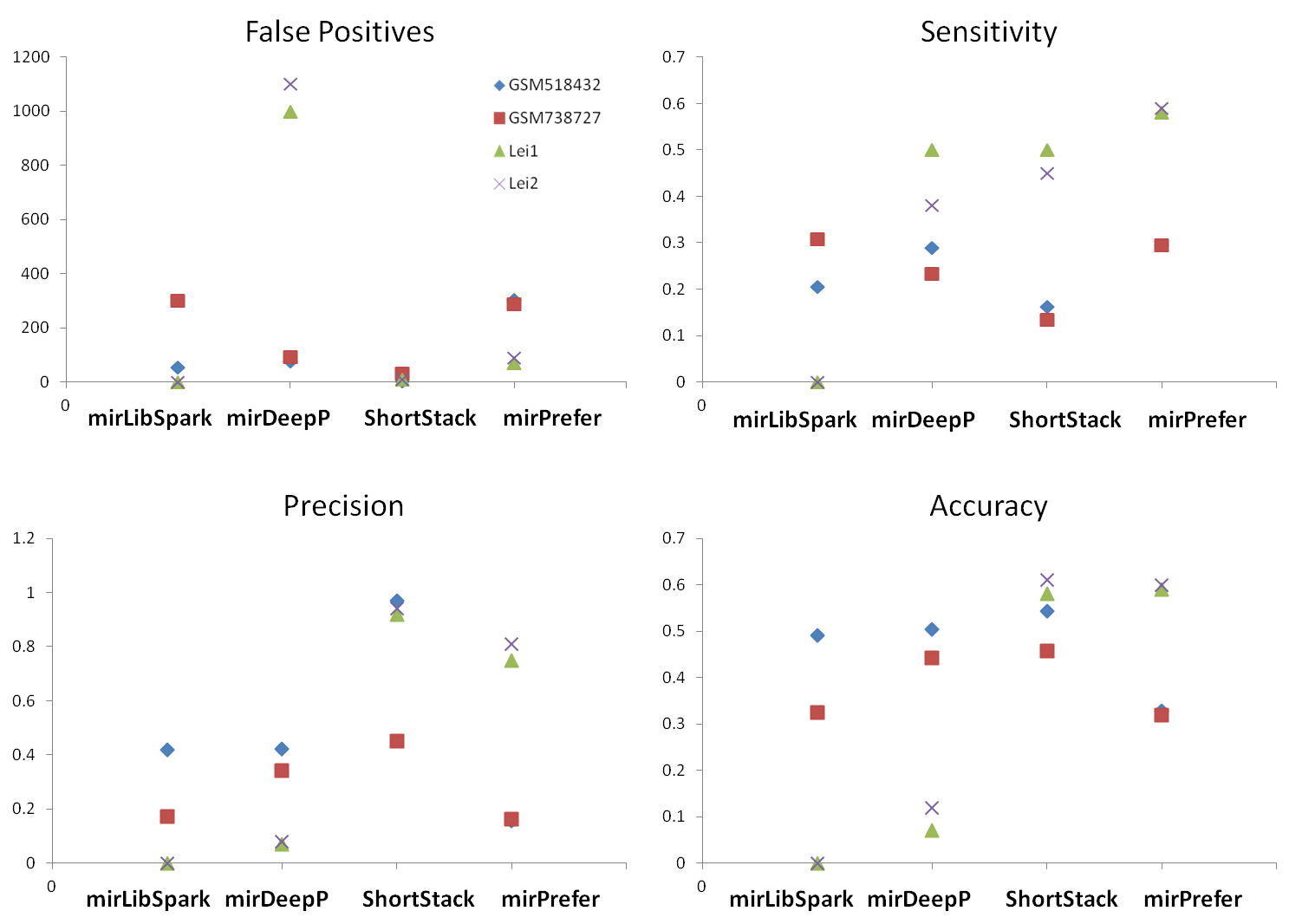}
\caption{Comparative performance of miRNA annotation software from plants.}
\label{FIG::Lei}
\end{figure}
\end{center}

We ran the programs of mirLibSpark, mirDeep-P, ShortStack and mirPrefer to process the libraries of GSM518432 and GSM738727 and obtained the miRNA predictions of each program, respectively.
Axtell and Meyer \citep{axtell2018revisiting} used a statistic method that is different than ours for Table \ref{TAB::compare_software} for assessing the performance.
In order to compare their assessments of plant annotation software with mirLibSpark,
we adapted their statistic method.
We thus defined the positive set as the miRBase-registered miRNAs (P).
The true positive predictions were the predicted miRNAs (TP), while the false negative predictions (FN) were the remaining instances in P.
The negative set (N) was defined as the miRBase-registered miRNAs not expressed in the library, and thus the same number across all programs.
The performance measures, including false positives, sensitivity, precision and accuracy, are shown in Figure \ref{FIG::Lei}.
We also overlaid the corresponding data points of the assessments by \citep{axtell2018revisiting} in our figures, labeled as \textit{Lei1} and \textit{Lei2}.
Because of the differences of the parameters used in the programs as well as the versions of the genome, RNAfold and so on, our assessments were not the same as theirs.

\subsection{Scalability}
MirLibSpark accelerates according to the number of core processors (nbCore).
As the nbCore increases, the time of execution decreases and the demand for memory space remains the same for the same input.
In addition, mirLibSpark is capable of processing large genome, such as wheat (see Table \ref{TAB::scalability}).
It reduced the time of processing of a wheat library from more than 30 hours by the non-parallelized program \textit{sequential.py} to 2.2 hours by mirLibSpark with 32 cores.

\begin{table}[h]
\centering
\begin{threeparttable}
\caption{Scalability of mirLibSpark for sRNA-seq data from small and large genomes.} 
\label{TAB::scalability}
\begin{tabular}{l|rrr}
\hline
(A) ath sRNA-seq lib    & nbCore & Time (sec) & MEM (GB)  \\ \hline
sequential.py           & 1      & 12396      & 0.8       \\ \hline
run\_mirdeep\_p.py      & 1      & 47752      & 0.7       \\ \hline
mirLibSpark             & 1      & 2980       & 14.1      \\
                        & 2      & 1472       & 13.2      \\
                        & 3      & 1071      & 13.6       \\
                        & 4      & 837       & 12.0       \\
                        & 8      & 501       & 12.3       \\
                        & 16     & 270       & 11.4       \\
                        & 32     & 215       & 10.2       \\ 
\hline\hline
(B) wheat sRNA-seq lib    & nbCore & Time (sec)  & MEM (GB)  \\ \hline
sequential.py             & 1      & > 30 hrs *  & 15.3      \\ \hline
mirLibSpark               & 8      & 23561       & 60.3      \\
                          & 16     & 13476       & 63.5      \\
                          & 32     & 8057        & 62.0      \\ 
\hline
\end{tabular}
\begin{tablenotes}
      \small
      \item The input library for panel (A) is \textit{Arabidopsis} GSM1087974 of 49 MB with genome 117 MB; 
      for panel (B) is \textit{T. aestivum} S002B35 of 2.1 GB with genome 13 GB. 
      *: The program was allocated 30 hours due to limited resources and terminated before reaching the end of the program.
      The results of time and memory are the average of three executions of the same settings. 
      Done in Graham computing cluster of Compute Canada with Intel E5-2683 V4 CPUs, running at 2.1 GHz and SATA SSD.
    \end{tablenotes}
  \end{threeparttable}
\end{table}

\subsection{Case study}
\begin{center}
\begin{figure}[h]
\centering
\includegraphics[width=8.5cm]{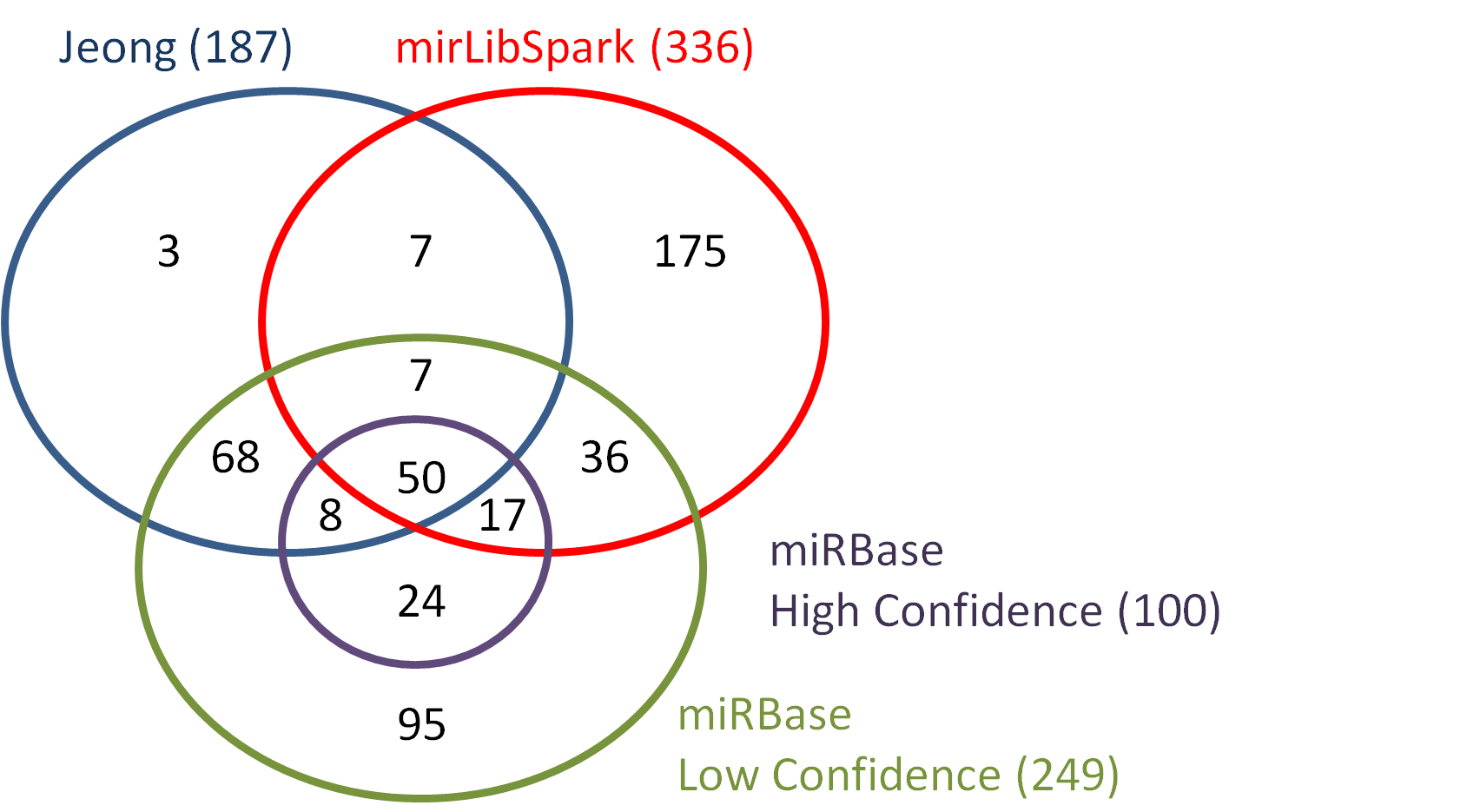}
\caption{The venn diagram characterizing the predictions of GSE44622 by Jeong et al \citep{jeong2013comprehensive} and mirLibSpark, after a clustering step using blastn to group together similar sequences.}
\label{FIG::case_VennDiagram}
\end{figure}
\end{center}

The miRNA prediction was reproduced for 27 \textit{Arabidopsis} sRNA librareis in GEO Series GSE44622 using mirLibSpark.
The computation of differential expressions of the predicted miRNAs and the enriched KEGG pathways of their target genes was guided for twenty experiment-control pairs as defined in a \textit{diffguide\_file}.
The entire process, including both prediction and functional annotation, took about 2 hours on Compute Canada Graham computing cluster (https://docs.computecanada.ca/wiki/Graham).
The miRNA prediction part alone took mirLibSpark 5,962 seconds, or 99 seconds per library on average.
After clustering the predicted miRNAs based on sequence similarity determined by blastn, the clusters of mirLibSpark prediction were compared with those clusters of the original paper and the miRBase-v21 high/low confidence, using venn digram as shown in Figure \ref{FIG::case_VennDiagram}.

Our results show high degrees of overlapping predictions with Jeong et al \citep{jeong2013comprehensive} and the known miRNAs.
Based on our simulation results, the non-overlapping clusters predicted by mirLibSpark (175 clusters) are possibly valid novel miRNAs.
However, they used the old version of the prediction criteria \citep{meyers2008criteria} while we added the updated criteria \citep{axtell2018revisiting}.
They removed some predicted miRNAs manually according to additional criteria that were not performed by mirLibSpark.
For example, if the sum of abundance in the 27 libraries is less than 400, or the genomic loci of the precursors are in siRNA-like regions, the miRNAs are removed \citep{jeong2013comprehensive}.
Although we filtered siRNAs using predefined genomic annotations, we did not identify genomic regions that are likely siRNA precursors.
Therefore, among our novel miRNAs collections, a fraction of functionally irrelevant predictions might not be excluded based on current criteria.

\section{Conclusion}
We facilitate the settings and annotations of several procedures, 
including the retrieval of the reference genome, 
building genomic indices,
and installing dependency programs and libraries.
MirLibSpark takes different input formats.
Most of the validation parameters are flexible in order to meet species-specific preferences or criteria updates.
This will allow a quick deployment and ready to use environment.
The user manuals for Docker and computing cluster environments, respectively, are available in GitHub.

We introduced the simulated data sets to properly access the performance with statistical measures.
We showed an improved accuracy on the prediction.
It is scalable regarding multiple libraries and libraries from large genomes.
The case study showed high degrees of overlapping with the prediction by others. 
One can easily reproduce results among research teams applying the stand-alone, dependency-self-supporting, and fully automated mirLibSpark.

In the future, all the added dependencies will be updated regularly.
The miRNAs predicted using mirLibSpark in wheat and other plant libraries will be collected in Wheat MicroRNA Portal \citep{remita2016novel}.
We will also investigate how to standardize the exclusion procedures in real-world data, such as identifying the miRNA precursors residing in the putative siRNA genomic regions and the low occurrences among libraries.

\bibliographystyle{plainnat}
\bibliography{main}

\begin{thebibliography}{29}
\providecommand{\natexlab}[1]{#1}
\providecommand{\url}[1]{\texttt{#1}}
\expandafter\ifx\csname urlstyle\endcsname\relax
  \providecommand{\doi}[1]{doi: #1}\else
  \providecommand{\doi}{doi: \begingroup \urlstyle{rm}\Url}\fi

\bibitem[Agharbaoui et~al.(2015)Agharbaoui, Leclercq, Remita, Badawi, Lord, Houde, Danyluk, Diallo, and Sarhan]{agharbaoui2015integrative}
Zahra Agharbaoui, Mickael Leclercq, Mohamed~Amine Remita, Mohamed~A Badawi, Etienne Lord, Mario Houde, Jean Danyluk, Abdoulaye~Banir{\'e} Diallo, and Fathey Sarhan.
\newblock An integrative approach to identify hexaploid wheat mirnaome associated with development and tolerance to abiotic stress.
\newblock \emph{BMC genomics}, 16\penalty0 (1):\penalty0 339, 2015.

\bibitem[Altschul et~al.(1997)Altschul, Madden, Sch{\"a}ffer, Zhang, Zhang, Miller, and Lipman]{altschul1997gapped}
Stephen~F Altschul, Thomas~L Madden, Alejandro~A Sch{\"a}ffer, Jinghui Zhang, Zheng Zhang, Webb Miller, and David~J Lipman.
\newblock Gapped blast and psi-blast: a new generation of protein database search programs.
\newblock \emph{Nucleic acids research}, 25\penalty0 (17):\penalty0 3389--3402, 1997.

\bibitem[Axtell(2013)]{axtell2013shortstack}
Michael~J Axtell.
\newblock Shortstack: comprehensive annotation and quantification of small rna genes.
\newblock \emph{Rna}, 2013.

\bibitem[Axtell and Meyers(2018)]{axtell2018revisiting}
Michael~J Axtell and Blake~C Meyers.
\newblock Revisiting criteria for plant microrna annotation in the era of big data.
\newblock \emph{The Plant Cell}, 30\penalty0 (2):\penalty0 272--284, 2018.

\bibitem[Benjamini and Hochberg(1995)]{benjamini1995controlling}
Yoav Benjamini and Yosef Hochberg.
\newblock Controlling the false discovery rate: a practical and powerful approach to multiple testing.
\newblock \emph{Journal of the royal statistical society. Series B (Methodological)}, pages 289--300, 1995.

\bibitem[Biegert et~al.(2006)Biegert, Mayer, Remmert, S{\"o}ding, and Lupas]{biegert2006mpi}
Andreas Biegert, Christian Mayer, Michael Remmert, Johannes S{\"o}ding, and Andrei~N Lupas.
\newblock The mpi bioinformatics toolkit for protein sequence analysis.
\newblock \emph{Nucleic acids research}, 34\penalty0 (suppl\_2):\penalty0 W335--W339, 2006.

\bibitem[Bolser et~al.(2017)Bolser, Staines, Perry, and Kersey]{bolser2017ensembl}
Dan~M Bolser, Daniel~M Staines, Emily Perry, and Paul~J Kersey.
\newblock Ensembl plants: integrating tools for visualizing, mining, and analyzing plant genomic data.
\newblock In \emph{Plant Genomics Databases}, pages 1--31. Springer, 2017.

\bibitem[Darty et~al.(2009)Darty, Denise, and Ponty]{darty2009varna}
K{\'e}vin Darty, Alain Denise, and Yann Ponty.
\newblock Varna: Interactive drawing and editing of the rna secondary structure.
\newblock \emph{Bioinformatics}, 25\penalty0 (15):\penalty0 1974, 2009.

\bibitem[de~Castro et~al.(2017)de~Castro, dos Santos~Tostes, D{\'a}vila, Senger, and da~Silva]{de2017sparkblast}
Marcelo~Rodrigo de~Castro, Catherine dos Santos~Tostes, Alberto~MR D{\'a}vila, Hermes Senger, and Fabricio~AB da~Silva.
\newblock Sparkblast: scalable blast processing using in-memory operations.
\newblock \emph{BMC bioinformatics}, 18\penalty0 (1):\penalty0 318, 2017.

\bibitem[Jeong et~al.(2013)Jeong, Thatcher, Brown, Zhai, Park, Rymarquis, Meyers, and Green]{jeong2013comprehensive}
Dong-Hoon Jeong, Shawn~R Thatcher, Rebecca~SH Brown, Jixian Zhai, Sunhee Park, Linda~A Rymarquis, Blake~C Meyers, and Pamela~J Green.
\newblock Comprehensive investigation of micrornas enhanced by analysis of sequence variants, expression patterns, argonaute loading, and target cleavage.
\newblock \emph{Plant physiology}, 162\penalty0 (3):\penalty0 1225--1245, 2013.

\bibitem[John et~al.(2004)John, Enright, Aravin, Tuschl, Sander, and Marks]{john2004human}
Bino John, Anton~J Enright, Alexei Aravin, Thomas Tuschl, Chris Sander, and Debora~S Marks.
\newblock Human microrna targets.
\newblock \emph{PLoS biology}, 2\penalty0 (11):\penalty0 e363, 2004.

\bibitem[Jones-Rhoades and Bartel(2004)]{jones2004}
Matthew~W Jones-Rhoades and David~P Bartel.
\newblock Computational identification of plant micrornas and their targets, including a stress-induced mirna.
\newblock \emph{Molecular cell}, 14\penalty0 (6):\penalty0 787--799, 2004.

\bibitem[Kal et~al.(1999)Kal, van Zonneveld, Benes, van~den Berg, Koerkamp, Albermann, Strack, Ruijter, Richter, Dujon, et~al.]{kal1999dynamics}
Arnoud~J Kal, Anton~Jan van Zonneveld, Vladimir Benes, Marlene van~den Berg, Marian~Groot Koerkamp, Kaj Albermann, Normann Strack, Jan~M Ruijter, Alexandra Richter, Bernard Dujon, et~al.
\newblock Dynamics of gene expression revealed by comparison of serial analysis of gene expression transcript profiles from yeast grown on two different carbon sources.
\newblock \emph{Molecular biology of the cell}, 10\penalty0 (6):\penalty0 1859--1872, 1999.

\bibitem[Kanehisa et~al.(2017)Kanehisa, Furumichi, Tanabe, Sato, and Morishima]{kanehisa2017kegg}
Minoru Kanehisa, Miho Furumichi, Mao Tanabe, Yoko Sato, and Kanae Morishima.
\newblock Kegg: new perspectives on genomes, pathways, diseases and drugs.
\newblock \emph{Nucleic Acids Research}, 45\penalty0 (D1):\penalty0 D353--D361, 2017.

\bibitem[Kanellos et~al.(2014)Kanellos, Vergoulis, Sacharidis, Dalamagas, Hatzigeorgiou, Sartzetakis, and Sellis]{kanellos2014mr}
Ilias Kanellos, Thanasis Vergoulis, Dimitris Sacharidis, Theodore Dalamagas, Artemis Hatzigeorgiou, Stelios Sartzetakis, and Timos Sellis.
\newblock Mr-microt: a mapreduce-based microrna target prediction method.
\newblock In \emph{Proceedings of the 26th International Conference on Scientific and Statistical Database Management}, page~47. ACM, 2014.

\bibitem[Kozomara et~al.(2018)Kozomara, Birgaoanu, and Griffiths-Jones]{kozomara2018mirbase}
Ana Kozomara, Maria Birgaoanu, and Sam Griffiths-Jones.
\newblock mirbase: from microrna sequences to function.
\newblock \emph{Nucleic acids research}, 47\penalty0 (D1):\penalty0 D155--D162, 2018.

\bibitem[Langmead et~al.(2009)Langmead, Trapnell, Pop, and Salzberg]{langmead2009ultrafast}
Ben Langmead, Cole Trapnell, Mihai Pop, and Steven~L Salzberg.
\newblock Ultrafast and memory-efficient alignment of short dna sequences to the human genome.
\newblock \emph{Genome biology}, 10\penalty0 (3):\penalty0 R25, 2009.

\bibitem[Lei and Sun(2014)]{lei2014mir}
Jikai Lei and Yanni Sun.
\newblock mir-prefer: an accurate, fast and easy-to-use plant mirna prediction tool using small rna-seq data.
\newblock \emph{Bioinformatics}, 30\penalty0 (19):\penalty0 2837--2839, 2014.

\bibitem[Li et~al.(2018)Li, Byrns, Badawi, Diallo, Danyluk, Sarhan, Laudencia-Chingcuanco, Zou, and Fowler]{li2018transcriptomic}
Qiang Li, Brook Byrns, Mohamed~A Badawi, Abdoulaye~Banire Diallo, Jean Danyluk, Fathey Sarhan, Debbie Laudencia-Chingcuanco, Jitao Zou, and D~Brian Fowler.
\newblock Transcriptomic insights into phenological development and cold tolerance of wheat grown in the field.
\newblock \emph{Plant physiology}, 176\penalty0 (3):\penalty0 2376--2394, 2018.

\bibitem[Lorenz et~al.(2011)Lorenz, Bernhart, H{\"o}ner~zu Siederdissen, Tafer, Flamm, Stadler, and Hofacker]{Lorenz2011}
Ronny Lorenz, Stephan~H. Bernhart, Christian H{\"o}ner~zu Siederdissen, Hakim Tafer, Christoph Flamm, Peter~F. Stadler, and Ivo~L. Hofacker.
\newblock Viennarna package 2.0.
\newblock \emph{Algorithms for Molecular Biology}, 6\penalty0 (1):\penalty0 26, Nov 2011.

\bibitem[McKenna et~al.(2010)McKenna, Hanna, Banks, Sivachenko, Cibulskis, Kernytsky, Garimella, Altshuler, Gabriel, Daly, et~al.]{mckenna2010genome}
Aaron McKenna, Matthew Hanna, Eric Banks, Andrey Sivachenko, Kristian Cibulskis, Andrew Kernytsky, Kiran Garimella, David Altshuler, Stacey Gabriel, Mark Daly, et~al.
\newblock The genome analysis toolkit: a mapreduce framework for analyzing next-generation dna sequencing data.
\newblock \emph{Genome research}, 2010.

\bibitem[Meyers et~al.(2008)Meyers, Axtell, Bartel, Bartel, Baulcombe, Bowman, Cao, Carrington, Chen, Green, et~al.]{meyers2008criteria}
Blake~C Meyers, Michael~J Axtell, Bonnie Bartel, David~P Bartel, David Baulcombe, John~L Bowman, Xiaofeng Cao, James~C Carrington, Xuemei Chen, Pamela~J Green, et~al.
\newblock Criteria for annotation of plant micrornas.
\newblock \emph{The Plant Cell}, 20\penalty0 (12):\penalty0 3186--3190, 2008.

\bibitem[Morgulis et~al.(2006)Morgulis, Gertz, Sch{\"a}ffer, and Agarwala]{morgulis2006fast}
Aleksandr Morgulis, E~Michael Gertz, Alejandro~A Sch{\"a}ffer, and Richa Agarwala.
\newblock A fast and symmetric dust implementation to mask low-complexity dna sequences.
\newblock \emph{Journal of Computational Biology}, 13\penalty0 (5):\penalty0 1028--1040, 2006.

\bibitem[Nellore et~al.(2016)Nellore, Collado-Torres, Jaffe, Alquicira-Hern{\'a}ndez, Wilks, Pritt, Morton, Leek, and Langmead]{nellore2016rail}
Abhinav Nellore, Leonardo Collado-Torres, Andrew~E Jaffe, Jos{\'e} Alquicira-Hern{\'a}ndez, Christopher Wilks, Jacob Pritt, James Morton, Jeffrey~T Leek, and Ben Langmead.
\newblock Rail-rna: Scalable analysis of rna-seq splicing and coverage.
\newblock \emph{Bioinformatics}, page btw575, 2016.

\bibitem[Pan et~al.(2015)Pan, Tseng, Liao, Chan, Lai, and Lin]{pan2015design}
Ren-Hao Pan, Lin-Yu Tseng, I-En Liao, Chien-Lung Chan, K~Robert Lai, and Kai-Biao Lin.
\newblock Design of an ngs microrna predictor using multilayer hierarchical mapreduce framework.
\newblock In \emph{Data Science and Advanced Analytics (DSAA). 36678. IEEE International Conference on}, pages 1--8. IEEE, 2015.

\bibitem[Remita et~al.(2016)Remita, Lord, Agharbaoui, Leclercq, Badawi, Sarhan, and Diallo]{remita2016novel}
Mohamed~Amine Remita, Etienne Lord, Zahra Agharbaoui, Mickael Leclercq, Mohamed~A Badawi, Fathey Sarhan, and Abdoulaye~Banir{\'e} Diallo.
\newblock A novel comprehensive wheat mirna database, including related bioinformatics software.
\newblock \emph{Current Plant Biology}, 7:\penalty0 31--33, 2016.

\bibitem[Vlachos et~al.(2015)Vlachos, Zagganas, Paraskevopoulou, Georgakilas, Karagkouni, Vergoulis, Dalamagas, and Hatzigeorgiou]{vlachos2015diana}
Ioannis~S Vlachos, Konstantinos Zagganas, Maria~D Paraskevopoulou, Georgios Georgakilas, Dimitra Karagkouni, Thanasis Vergoulis, Theodore Dalamagas, and Artemis~G Hatzigeorgiou.
\newblock Diana-mirpath v3. 0: deciphering microrna function with experimental support.
\newblock \emph{Nucleic acids research}, page gkv403, 2015.

\bibitem[Yang and Li(2011)]{yang2011mirdeep}
Xiaozeng Yang and Lei Li.
\newblock mirdeep-p: a computational tool for analyzing the microrna transcriptome in plants.
\newblock \emph{Bioinformatics}, 27\penalty0 (18):\penalty0 2614--2615, 2011.

\bibitem[Zaharia et~al.(2012)Zaharia, Chowdhury, Das, Dave, Ma, McCauley, Franklin, Shenker, and Stoica]{zaharia2012resilient}
Matei Zaharia, Mosharaf Chowdhury, Tathagata Das, Ankur Dave, Justin Ma, Murphy McCauley, Michael~J Franklin, Scott Shenker, and Ion Stoica.
\newblock Resilient distributed datasets: A fault-tolerant abstraction for in-memory cluster computing.
\newblock In \emph{Proceedings of the 9th USENIX conference on Networked Systems Design and Implementation}, pages 2--2. USENIX Association, 2012.

\end{thebibliography}
\end{document}